\documentclass[prb,superscriptaddress,twocolumn,preprintnumbers,a4,showpacs,floatfix]{revtex4}
\usepackage{amsmath, amssymb, psfrag, tabls, dcolumn, bm, color}
\usepackage[sort&compress]{natbib}
\usepackage[pdftex]{graphicx}
\usepackage{graphicx}
\usepackage{bm}

\newcommand{\be}{\begin{equation}}
\newcommand{\ee}{\end{equation}}
\newcommand{\bea}{\begin{equation*}}
\newcommand{\eea}{\end{equation*}}
\newcommand{\ba}{\begin{array}}
\newcommand{\ea}{\end{array}}
\newcommand{\beqa}{\begin{eqnarray}}
\newcommand{\eeqa}{\end{eqnarray}}
\newcommand{\beqaa}{\begin{eqnarray*}}
\newcommand{\eeqaa}{\end{eqnarray*}}

\newcommand{\matr}{\left( \begin{array}}
\newcommand{\ematr}{\end{array} \right)}

\newcommand{\der}{{\rm d}}

\newcommand{\lsim}{{\;\raise0.3ex\hbox{$<$\kern-0.75em\raise-1.1ex\hbox{$\sim$}}
\;}}
\newcommand{\gsim}{{\;\raise0.3ex\hbox{$>$\kern-0.75em\raise-1.1ex\hbox{$\sim$}}
\;}}

\begin{document}

\title{Twisting Graphene Nanoribbons into Carbon Nanotubes}

\author{O. O. Kit}
\affiliation{NanoScience Center, Department of Physics, University of Jyv\"askyl\"a, 40014 Jyv\"askyl\"a, Finland}

\author{T. Tallinen}
\affiliation{NanoScience Center, Department of Physics, University of Jyv\"askyl\"a, 40014 Jyv\"askyl\"a, Finland}
\affiliation{School of Engineering and Applied Sciences, Harvard University, 29 Oxford Street, Cambridge, MA 02138}

\author{L. Mahadevan}
\affiliation{School of Engineering and Applied Sciences, Harvard University, 29 Oxford Street, Cambridge, MA 02138}

\author{J. Timonen}
\affiliation{NanoScience Center, Department of Physics, University of Jyv\"askyl\"a, 40014 Jyv\"askyl\"a, Finland}

\author{P. Koskinen\footnote{Corresponding author}}
\email[email:]{pekka.koskinen@iki.fi}
\affiliation{NanoScience Center, Department of Physics, University of Jyv\"askyl\"a, 40014 Jyv\"askyl\"a, Finland}




\begin{abstract}
Although carbon nanotubes consist of honeycomb carbon, they have never been fabricated from graphene directly. Here we show, by quantum molecular-dynamics simulations and classical continuum-elasticity modeling, that graphene nanoribbons can, indeed, be transformed into carbon nanotubes by means of twisting. The chiralities of the tubes thus fabricated can be not only predicted but also externally controlled. This twisting route is a new opportunity for nano-fabrication, and is easily generalizable to ribbons made of other planar nanomaterials.
\end{abstract}

\maketitle

\section{Introduction}

Carbon nanotubes (CNTs) are today grown by atomic self-organization, resulting in variable tube diameters and structures\cite{iijima_nature_91,saito_book_98}, whereas in textbooks nanotubes are portrayed as being made by ``rolling-up graphene''. Recent experiments have shown how nanotubes can be unzipped into graphene nanoribbons (GNRs).\cite{kosynkin_nature_09,jiao_nature_09} The inverse route of making nanotubes of graphene directly, however, has remained elusive.

\begin{figure}[b!]
\includegraphics[width=7cm]{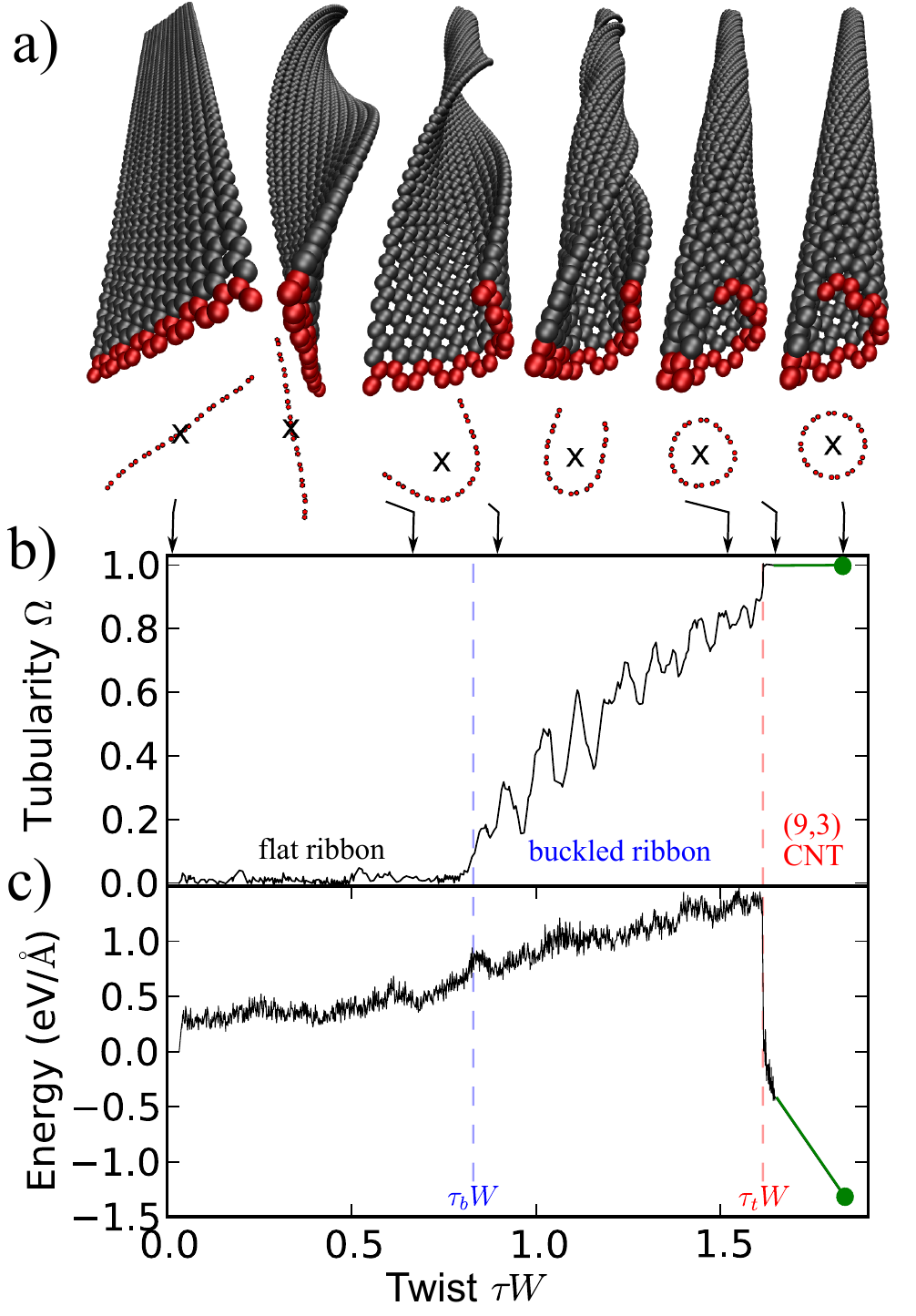}
\caption{(Color online) Quantum molecular-dynamics simulation of GNR transformation into a CNT. (a) The simulation of a $12$-ZGNR with snapshots at different twist parameters $\tau W$ (for GNR notations, see Appendix \ref{appendix:geometries}). The red atoms highlight one row of atoms across the ribbon, and their cross-section projections are shown below; the cross denotes the twist axis. The rightmost tube is the final, fully-optimized $(9,3)$ CNT. (b) The tubularity parameter $\Omega$ that shows the stages of flat ribbon, buckled ribbon, and CNT. Fluctuations in $\Omega$ arise from non-zero temperature. (c) Potential energy per unit length as a function of twist. The final curve segments in panels b and c (the green ones) represent the subsequent full structural optimization of the resulting CNT.}
\label{fig:demonstration}
\end{figure}

Here we show that graphene nanoribbons can indeed be transformed into carbon nanotubes by twisting. Our quantum molecular-dynamics simulations show that tube formation is preceded by buckling that serves as an intermediate stage between a flat ribbon and a pristine tube. Both buckling and tube formation can be explained by classical continuum elasticity, apart from effects due to atomic discreteness. Since graphene nanoribbons can be made with controlled edge morphology\cite{jia_science_09}, even with atomic precision and sub-nanometer width\cite{cai_nature_10},
this twisting route to carbon nanotube fabrication provides a qualitatively new opportunity for precision control in nanotube fabrication, while also enabling encapsulation of molecules, designed chirality, and easy generalization to ribbons made of other planar nanomaterials.

\section{The tube formation process: showcase}

To demonstrate the mechanism of tube formation, we first present a simulation of the twisting of a $24$~\AA\ wide, infinitely long graphene nanoribbon with unpassivated zigzag edges, shown in Fig.1a. We simulated the \textit{fixed-length} ribbon with quantum molecular-dynamics at room temperature, while increasing the twist at a constant rate; increasing the twist continuously without end effects was enabled by periodic boundary conditions adapted to chiral symmetry (see Appendix \ref{appendix:MD} for simulation details). The amount of twist is characterized by a dimensionless parameter $\tau W$ with $\tau$ being the twist angle per unit length and $W$ the width; hence $\tau W=0$ for a flat ribbon and $\tau W=1$ for a ribbon that has been twisted a full turn within a length of $2\pi W$.

At the initial phases of the simulation, the cross section of the ribbon remained flat until a critical value $\tau W=\tau_b W\cong 0.83$ when the ribbon buckled into a twisted groove with a U-shaped cross section, see Fig.1a. In order to characterize the tubular geometry quantitatively, we introduced a geometrical parameter, the tubularity parameter $\Omega$, which measures how much the ribbon edges have approached each other in relation to ribbon width: it is zero for flat ribbons and one for tubes [see Eq.(\ref{eq:omega})].

Upon further twisting, ribbon's tubularity was fluctuating, but increased on the average. Increased twisting brought therefore the opposite edges closer together, and at a second critical value, $\tau W=\tau_t W\cong 1.62$, they began to interact chemically: they were joined by a sudden formation of bonds, and the buckled ribbon was rapidly transformed into a tube. This point was identified as a sudden increase in the tubularity parameter $\Omega$ and as an even more sudden decrease in the potential energy due to the formation of $\sigma$ and $\pi$ bonds. At this stage, the resulting tube had a slightly non-round cross section. So as to anneal any remnant residual strains, we stopped the molecular-dynamics simulation and carried out a full structural optimization. The simulation then resulted in a pristine $(9,3)$ CNT (see Supplemental Video 1).\cite{hamada_PRL_92} We repeated such simulations for a number of zigzag and armchair nanoribbons of varying width, and invariably obtained pristine CNTs.

\begin{figure}[b]
\includegraphics[width=8cm]{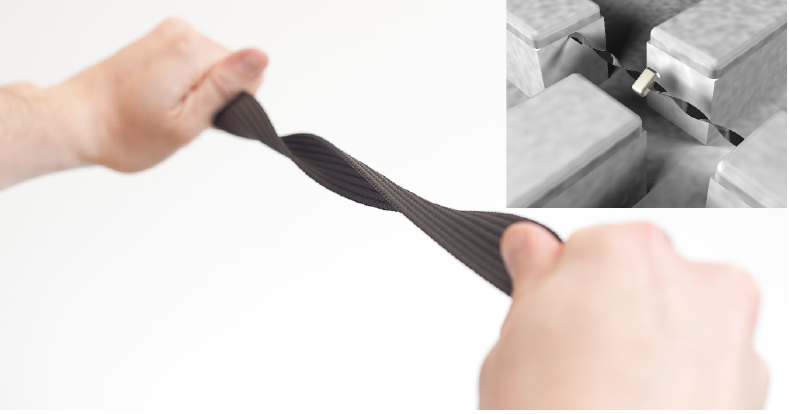}
\caption{(Color online) Ribbon buckling and tube formation can be demonstrated with a strap of a backpack. Inset: Proposal for an experimental realization, adapting the successful scheme of
Refs~\onlinecite{fennimore_nature_03} and \onlinecite{meyer_science_05}. }
\label{fig:strap}
\end{figure}

The essence of the tube formation process outlined above can be captured by a simple demonstration---just twist the ends of a strap of your backpack and watch the result (Fig.2). In particular, the CNT formation process is not an artifact of the imposed boundary conditions; simulations with finite tubes and thousands of atoms produce the same results, as discussed in Section \ref{sec:pbc}. On the actual nano-scale, twisting experiments should be feasible using the established paddle-type setups in which voltages applied to electrodes can be used to control the orientation of the somewhat asymmetrically positioned paddle electrostatically (the inset in Fig.2a). This setup has already been successful in twisting CNTs.\cite{fennimore_nature_03,meyer_science_05}

These atomistic simulations showed the feasibility of tube formation by twisting, but raised several questions that include characterization of buckling and tube formation as a function of ribbon width, the required critical torque, the resulting CNT chiralities and their possible control, electronic structure modifications, finite-size effects, effects of imperfections, and, finally, a simple explanation of the mechanisms involved at a classical continuum level. We attempt to address these questions in the following sections.

\section{Tube formation is governed by continuum elasticity}

\begin{figure*}
\includegraphics[width=14cm]{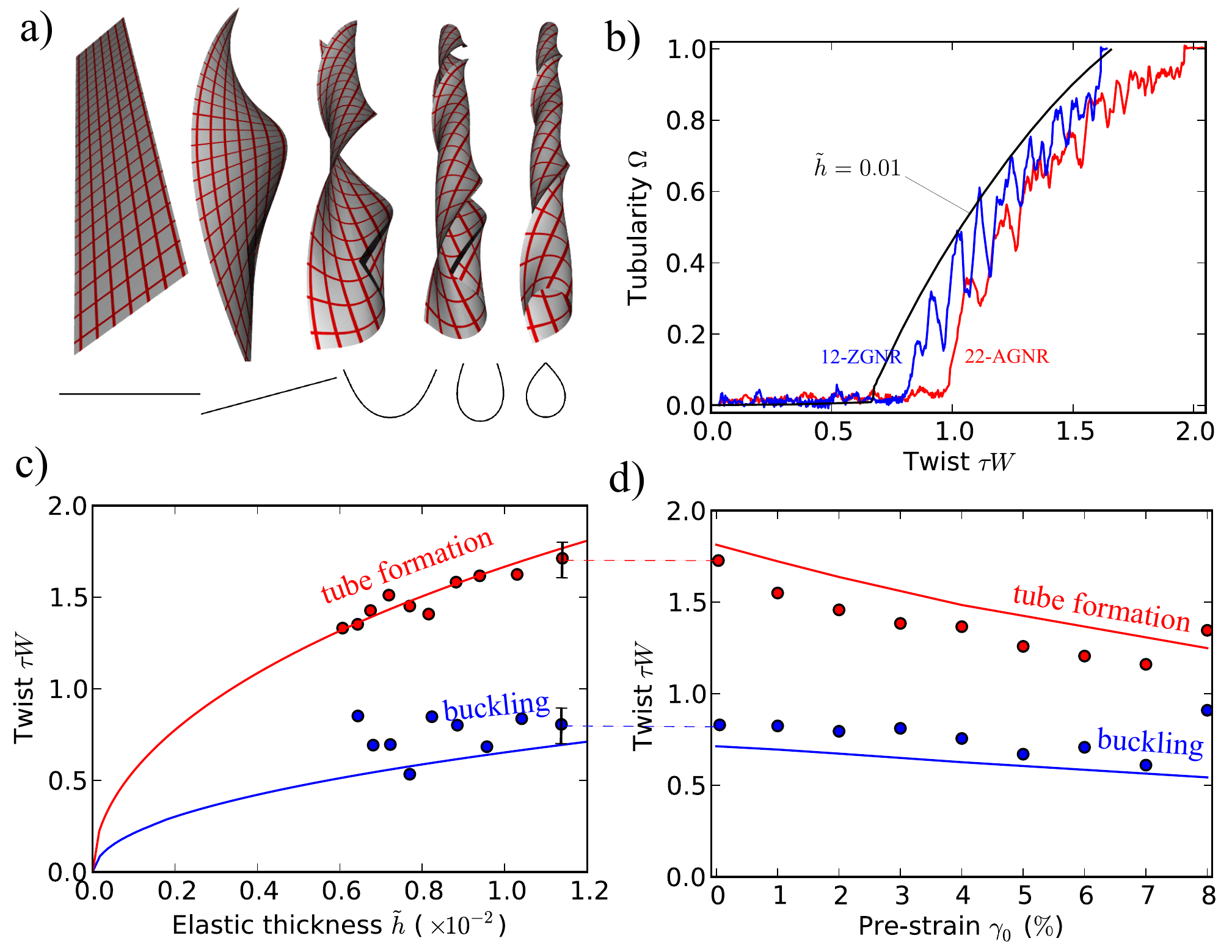}
\caption{(Color online) Tube formation by elasticity theory. (a) Visualizations of twisted elastic ribbons for $\tilde{h}=0.01$ (width $\approx 24$~\AA\ in graphene) and $\tau W=0$, $0.44$, $0.88$, $1.32$, and $1.65$. Cross-section projections  are also shown. (b) Tubularity parameter $\Omega$ by elasticity theory ($\tilde{h}=0.01$) and from simulation results for a zigzag ($12$-ZGNR, $\tilde{h}\approx 0.0104$) and an armchair ribbon ($22$-AGNR, $\tilde{h}\approx 0.0097$). (c) Buckling ($\tau_b$) and tube-formation points ($\tau_t$) as a function of elastic thickness, which result from elasticity theory and from simulations for different ZGNRs. For the rightmost ribbon, $10$-ZGNR, the simulation was repeated $10$ times, and the vertical bars denote the range where $\tau_b$ and $\tau_t$ were observed to fluctuate. (d) Buckling and tube-formation points that result from elasticity theory ($\tilde{h}=0.012$) and from simulation results for $10$-ZGNR ($\tilde{h}\approx 0.0117$), as a function of pre-strain.}
\label{fig:elasticity}
\end{figure*}

We addressed the last question by modeling GNRs as thin elastic sheets with an in-plane modulus $k$, bending modulus $K$, Poisson ratio $\nu=0.3$, and width $W$. The twisted shape was found by numerically minimizing the elastic deformation energy that had an in-plane stretching component and an out-of-plane bending component (see Appendix \ref{appendix:elasticity}). Our analysis based on continuum elasticity theory indicates that the key elements in tube formation are the following: Buckling results from transverse stress in the twisted ribbon and the shape of the buckled cross section from competition between stretching and bending.

The condition of fixed length in the analysis distinguishes it from previous studies in which the tensile force was fixed, and ribbon length could thus vary leading to smooth longitudinal buckling~\cite{green_PRSL_36,green_PRSL_37}, helical developable geometry~\cite{mansfield_book_89}, or triangular stress-focusing patterns \cite{korte_PRSL_10}. Our analysis assumed that the cross section of the ribbon was free to bend and warp, which is not possible close to its ends if they are clamped. The ribbon should thus be sufficiently long for our assumptions to be valid. The length $2 \pi/\tau_t$, over which the ribbon makes a full turn at the point of tube formation, can be taken as a reasonable minimum-length criterion.

Dimensional analysis indicates that the twisted geometry depends only on the dimensionless elastic thickness, $\tilde{h} = \sqrt{K/(kW^2)}$, and twist, $\tau W$. For graphene $K = 1.6$ eV and $k = 25$ eV/\AA$^2$ such that $\tilde{h} \approx 0.01$ in the above simulation with $W \approx 24$~\AA.\cite{koskinen_PRB_10b} Figure~\ref{fig:elasticity}b shows the changing geometry of an elastic ribbon under twisting for $\tilde{h} = 0.01$---the similarity with Fig.~\ref{fig:demonstration}a is evident. Quantitative comparison in Fig.~\ref{fig:elasticity}b of the tubularity parameters of quantum and classical analyses makes the similarity of these two approaches even more evident, although there are some differences near $\Omega \approx 1$. This is because of the difference between the zigzag and armchair ribbons of similar width, which have somewhat different tube-formation mechanisms due to the role of edge morphology and edge stress \cite{liang_PNAS_09,bhuang_PRL_09} that do not normally appear in continuum models (discussed in Section \ref{sec:armchair}). We note here that the  maximum strain induced by twisting at the ribbon edge  is $\approx 6\%$ for $\tilde{h} = 0.01$, decreasing as $\sim 1.3\text{ \AA}/W$ for increasing $W$, so graphene ribbons survive twisting without tearing.

The buckling point, $\tau_b$, and the tube formation point, $\tau_t$, vary with the scaled elastic thickness, $\tilde{h}$, as shown in Fig.~\ref{fig:elasticity}c. These numerically obtained curves are well approximated by the expressions $\tau_b W \approx 6.6 \sqrt{\tilde{h}}$ and $\tau_t W \approx 17.2 \sqrt{\tilde{h}}$ that follow from a simple scaling analysis (see Appendix \ref{appendix:elasticity}). The same analysis also yields the torque associated with the tube-formation point as $M_t \approx 1.8 K/\sqrt{\tilde{h}}$. In the presence of an externally applied pre-strain, $\gamma_0$, which naturally can be released after tube formation, the critical values for $\tau_b$ and $\tau_t$ decrease as shown in Fig.~\ref{fig:elasticity}d. This is in agreement with the trend observed in molecular-dynamics simulations: Control over $\gamma_0$ thus provides a way to fine-tune the required twist and has implications for the resulting CNT chiralities, which is the question we discuss next.

\section{CNT Chiralities can be predicted}
\def\deltaa{\delta^i}
\def\deltab{\delta^h}

\begin{figure*}
\includegraphics[width=13cm]{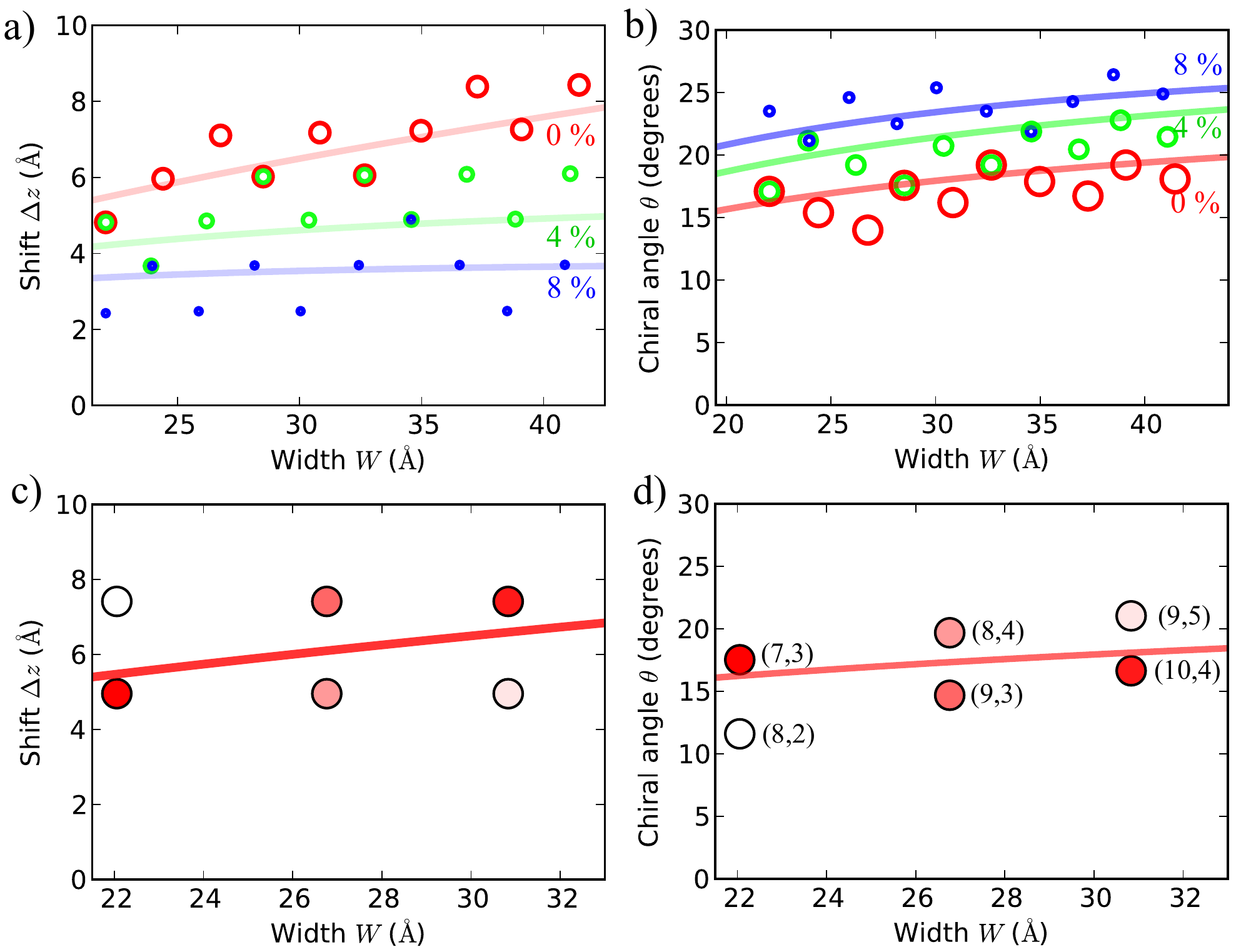}
\caption{(Color online) Predicting CNT chiralities. (a) Axial shift $\Delta z$ of the opposite edges in the tube as given by the elasticity theory as a function of ribbon width (lines) for three values of pre-strain (denoted by percentages). Circles are the simulated values for ZGNRs. (b) The CNT chiral angles, $\theta$, that correspond to the shifts in panel a. Full lines are the predictions for $\theta$ given by Eq.(\ref{eq:apprchirality}) with $\phi=0^\circ$ and with $\Delta z/W$ given by the elasticity theory. (c) Shift $\Delta z$ for $10-$, $12-$ and $14-$ZGNRs (symbols from left to right) and the corresponding elasticity-theory prediction (full curve) with zero pre-strain. Each of these ribbons was simulated $10$ times, and the intensity of the color inside the circles is proportional to the number of times the room-temperature simulation gave the given $\Delta z$. (d) The CNT chiral angles, augmented by the chiral indices, corresponding to the $\Delta z$ of c, with the same definitions for the symbols. The line is the elasticity-theory prediction by Eq.(\ref{eq:apprchirality}). }
\label{fig:chirality}
\end{figure*}

Our simulations suggest that, by knowing the width and chirality of the GNR, the chirality of the CNT can be predicted with unexpected reliability. The CNT chiral angle, $\theta$, depends on a shift, $\Delta z$, that measures the (relative) axial displacement of ribbon edges at the tube-formation point. In the continuum picture the honeycomb geometry thus implies the relation
\begin{equation}
 \phi + \theta = \pi/2 - \arctan (\Delta z/W)
\label{eq:apprchirality}
\end{equation}
between the GNR ($\phi$) and CNT ($\theta$) chiral angles\cite{barone_NL_06,saito_book_98} that vary between $0^\circ$ and $30^\circ$ (see Appendix \ref{appendix:geometries}). With $\Delta z/W$ given by the elasticity theory and with tube circumference deduced from the ribbon width (Fig.~\ref{fig:diameter}), Eq.(\ref{eq:apprchirality}) offers a recipe for a continuum prediction of the CNT chirality.

In the atomistic picture, however, the shift $\Delta z$ has to be compatible with the atomic discreteness. Figure~\ref{fig:chirality}a shows how $\Delta z$ in $N$-ZGNRs picked values close to the continuum predictions, either an integer or a half-odd-integer times the edge periodicity $a_\text{zz}=2.46$~\AA. Other values of $\Delta z$ would have cost additional shear-deformation energy. Similarly, Fig.~\ref{fig:chirality}b shows that the chiral angles of the corresponding CNTs' pick allowed values in the proximity of the continuum-limit curves given by Eq.(\ref{eq:apprchirality}).

The elasticity result for $\Delta z$ can, in fact, be used for a still better estimate of the CNT chiral angle---also for an estimate of the CNT chiral indices $(n,m)$ themselves. For a given $N$ and ribbon type, only certain indices $(n,m)$ are allowed, and the honeycomb lattice suggests the expressions
\begin{equation}
\label{eq:chirality}
(n,m) =
\begin{cases}
(N/2+\deltab, N/2-\deltab),&\mbox{\hspace{-0.3cm}odd-}N\mbox{ ZGNR}\\
(N/2+\deltaa, N/2-\deltaa),&\mbox{\hspace{-0.3cm}even-}N\mbox{ ZGNR}\\
(N/2-\deltab, 2 \deltab),&\mbox{\hspace{-0.3cm}odd-}N\mbox{ AGNR}\\
(N/2-\deltaa, 2 \deltaa),&\mbox{\hspace{-0.3cm}even-}N\mbox{ AGNR},
\end{cases}
\end{equation}
where $\deltaa$ is the nearest integer and $\deltab$ is the nearest half-odd-integer to the ratio $\Delta z/a_{[\text{zz}/\text{ac}]}$, owing to the symmetry of the opposite edges as shown in Fig.~\ref{fig:ribbons} ($N/2$ can be a half-odd-integer), and where $\Delta z$ is predicted by the elasticity theory. Prediction for the CNT chiral angle is then given by $(n,m)$ via the expression
\begin{equation}
\theta(n,m) = \arccos \frac{2 n+m}{2\sqrt{m^2+n^2+mn}}.
\end{equation}

\begin{figure*}
\includegraphics[width=17cm]{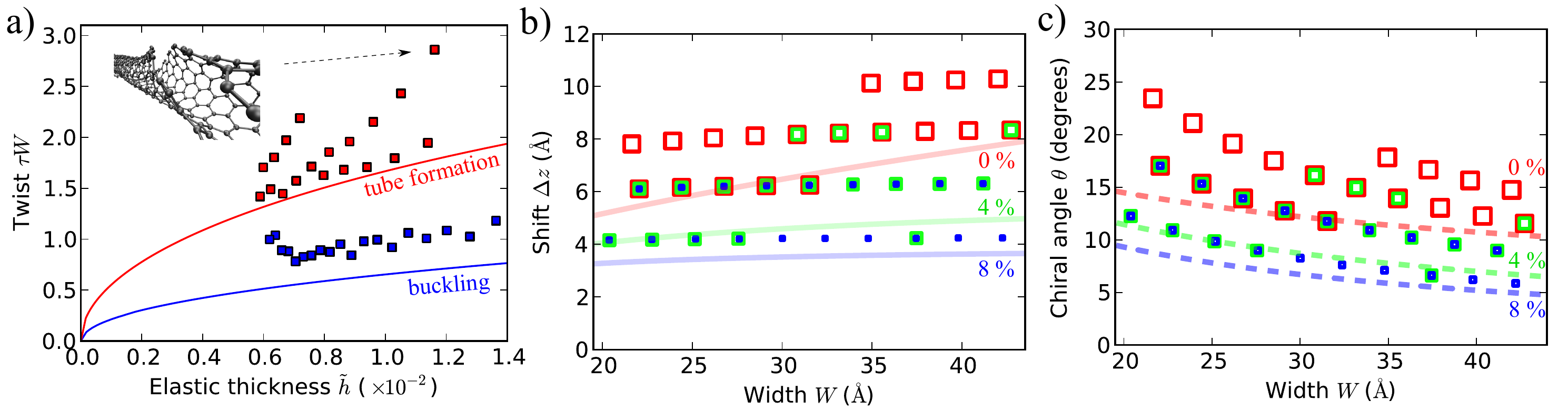}
\caption{(Color online) Tube formation for armchair graphene nanoribbons. (a) Buckling ($\tau_b$) and tube-formation points ($\tau_t$) as a function of elastic thickness given by the elasticity theory (full lines; Eq.(\ref{eq:apprchirality}) with $\phi=30^\circ$) and by simulations (squares) for AGNRs of varying width (corresponds to Fig.~\ref{fig:elasticity}c for ZGNRs). The inset: AGNR prior to tube formation, illustrating the radial bulging of the edges. Pre-strain is zero. b) Shift $\Delta z$ as given by the elasticity theory (full lines) and by atomic simulations (squares) as a function of ribbon width for three values of pre-strain (corresponds to Fig.~\ref{fig:chirality}a for AGNRs) (c) The CNT chiral angles corresponding to the shifts in b.}
\label{fig:armchair}
\end{figure*}

Yet Eq.(\ref{eq:chirality}) is only a prediction---the tube formation process contains stochastic aspects. For some ribbons the continuum-limit value for $\Delta z$ happened to be halfway between two allowed values, making prediction based on room-temperature simulations inevitably less precise. For example, Fig.~\ref{fig:chirality}c and \ref{fig:chirality}d show how identical room temperature simulations result in different shifts $\Delta z$ and in correspondingly different CNT chiralities. While the obtained shifts indeed favored the nearest integer multiple of $a_\text{zz}$ close to the elasticity-theory prediction, they fluctuated so that the final result could not be predicted with certainty. Still for more than $80$~\%\ of our simulations Eq.(\ref{eq:chirality}) predicted the CNT chirality correctly.

Finally, since increasing the axial pre-strain $\gamma_0$ decreases $\tau_t$, and thus $\Delta z$, as shown in Fig.~\ref{fig:chirality}a, we can use $\gamma_0$ to control the CNT chirality as illustrated in Fig.~\ref{fig:chirality}b. In particular, given the discreteness of the allowed $\Delta z$, even a coarse experimental control over $\gamma_0$ can be used to fine-tune the chirality.

\section{Armchair ribbons: Edge stress and forced-joining effects}
\label{sec:armchair}

Figure~\ref{fig:armchair}a shows the buckling and tube-formation points for AGNRs, showing less apparent agreement between simulations and continuum-limit theory than for ZGNRs (compare with Fig.~\ref{fig:elasticity}c). These differences have two reasons: compressive edge stress and `forced-joining effect' that arises from the comparatively large unit-cell length at the ribbon edge.

The compressive stress at the edge of an unpassivated AGNRs is $\approx 1.45$~eV/\AA, some $\approx 3.5$ times greater than in ZGNRs.\cite{bhuang_PRL_09} In narrow ribbons this stress can cause spontaneous twisting because of elongation of the edge with respect to the ribbon axis.\cite{koskinen_PRL_10} Because the stress makes edges to prefer small strain, larger twists were required for buckling (consistent shifts between the blue curve and the blue symbols in Fig.~\ref{fig:armchair}a). However, we omitted the edge stress from the elasticity theory on three grounds. First, it affected mainly the buckling threshold. Second, the edge stress depends on edge type and passivation---with hydrogen passivation the stress vanishes. Third, the edge stress can be imitated simply by having a decreased pre-strain.

Apart from buckling, the tube formation itself was dominated by a kind of forced-joining effect caused by edge morphology. Namely, because the edge periodicity $a_\text{ac}=4.26$~\AA\ in AGNRs is almost twice the periodicity $a_\text{zz}=2.46$~\AA\ in ZGNRs, larger deviations from the continuum prediction for $\Delta z$ were required---certain `forcing' was needed to initiate the tube formation. Joining was easier for wide ribbons, where shifts in steps of $a_\text{ac}$ required less shear, and for ZGNRs, where the required steps $a_\text{zz}$ were smaller.

Furthermore, prior to tube formation the edge stress turned our to make the edges of buckled ribbons to bulge radially outwards (the inset of Fig.~\ref{fig:armchair}a). Then, to initiate the tube formation, the buckled ribbon with bulged edges often required more forced twisting so as to attain the allowed $\Delta z$. In some simulations bulging was further enhanced by twisting, making tube formation to require exceptionally large twists (tube formation was hindered by an energy barrier). In these simulations a longer simulation time or higher temperature might have initiated the tube formation earlier. However, when the continuum $\Delta z$ happened to be such that dangling bonds from the opposite edges met directly, no forcing was needed and simulation yielded $\tau_t$ in agreement with the elasticity theory prediction (compare Figs~\ref{fig:armchair}a and \ref{fig:armchair}b).

Note that while pre-strain drives ZGNRs towards armchair CNTs ($\theta=30^\circ$; see Fig.~\ref{fig:chirality}b), it drives AGNRs towards zigzag CNTs ($\theta=0^\circ$; see Fig.~\ref{fig:armchair}b). These tendencies can be understood by noticing that increased pre-strain always decreases $\Delta z$; with precisely zero $\Delta z$ armchair ribbons would become zigzag tubes and zigzag ribbons armchair tubes.

\begin{figure}[b]
\includegraphics[width=6.5cm]{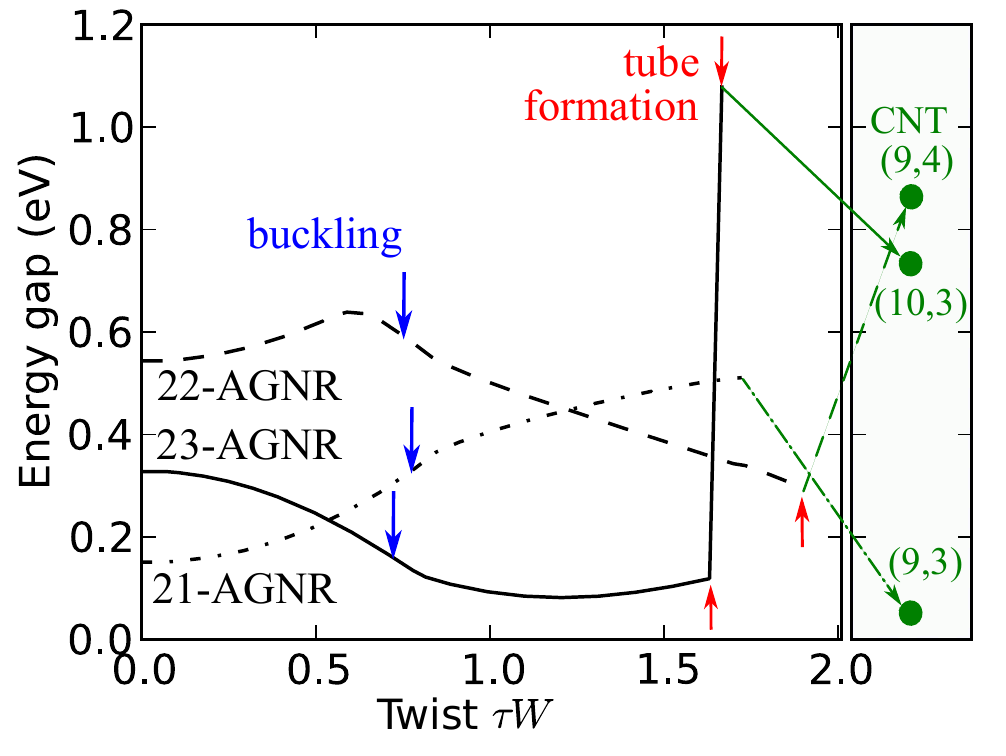}
\caption{(Color online) Energy gaps for selected $N$-AGNRs from quasi-static simulations (optimized atomic coordinates for given $\tau W$), representing three different $N$ families. The narrow panel shows the gaps in the resulting CNTs with torque and axial stress removed.}
\label{fig:gaps}
\end{figure}

\begin{figure*}
\includegraphics[width=15cm]{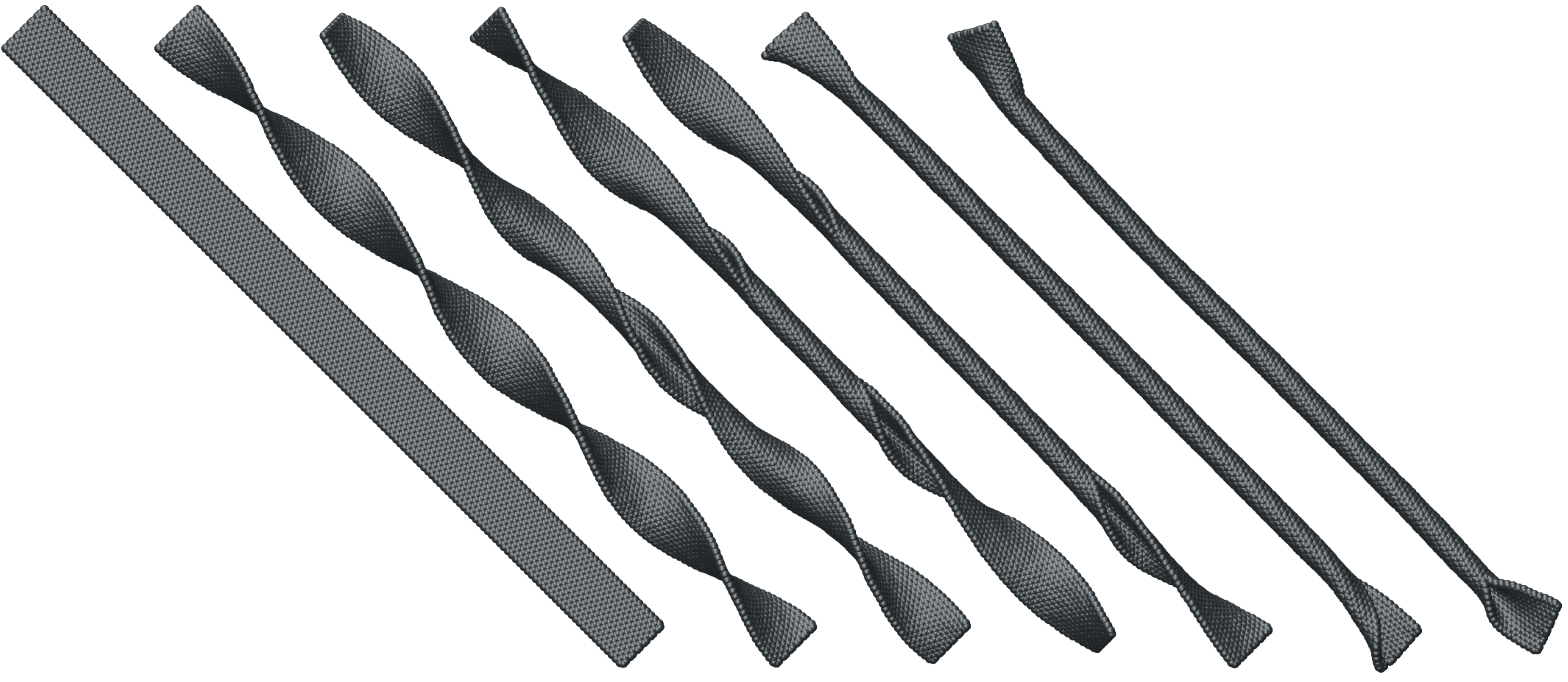}
\caption{(Color online) Finite ribbon simulations. Figure shows snapshots from the simulation of $L=38$~nm long $12$-ZGNR with clamped ends and with increasing end-to-end twist angle (the first six snapshots; structures are rotated to give the best view angles). The last snapshot is the (partially unzipped) $(9,3)$ CNT after removing the clamps and performing a constraint-free simulation at $300$~K.}
\label{fig:finite}
\end{figure*}

In addition to the purely geometrical effects associated with the GNR-CNT transition, we also investigated twisting-induced modifications in the electronic properties of the system. Prior to buckling, the energy gaps in $N$-AGNRs were observed to form three ``families'' according to $q=\text{mod}(N,3)$, as reported earlier\cite{gunlycke_NL_10,zhang_small_11,koskinen_APL_11}, and as illustrated in Fig.~\ref{fig:gaps}. Buckling turned out to cause only slight rehybridization of the carbon atoms, consistent with earlier studies on CNTs.\cite{popov_NJP_04} In contrast to the smooth electronic modification caused by twisting, at the point of tube formation the gap instantaneously jumped to that of CNT---which may have been, prior to relaxation, still influenced by remnant residual torques.

\section{Not an artifact of \newline periodic boundary conditions}
\label{sec:pbc}

The validity of the revised periodic boundary condition (RPBC) approach was confirmed by finite-ribbon simulations. Here we present exemplary results of a simulation for $L=38$~nm long $12$-ZGNR with $3720$ atoms. Ribbon's one end was clamped, and the other end was kept at a fixed distance (without pre-strain) while twisted continuously at the rate $\der \theta /\der t=0.0091$~degrees/fs with a $2$~fs time step; the rest of the atoms were treated with a Langevin thermostat set to $300$~K. In this simulation we used the REBO interatomic potential from the LAMMPS package.\cite{plimpton_JCP_95,lammps,brenner_JPCM_02}

Supplementary Video 2 shows an animation of this simulation, with selected snapshots in Fig.~\ref{fig:finite}. The tube-formation processes in RPBC and in finite ribbons were the same---and both resulted in the same pristine $(9,3)$ CNT. Sure enough, some finite-size effects did arise. Both buckling and tube formation initiated in a narrow central region, and the zipping-up propagated towards the ribbon ends when the applied twist increased. Complex distortions were suppressed by the experimentally feasible fixed-length constraint. After tube formation we released the end constraints and observed that the tubes (that were partially unzipped near the ends) remained thermodynamically stable at $300$~K, and even at $1200$~K.

The spreading of the end-to-end twist angle $\Delta \theta$ across the ribbon was somewhat uneven, and twisting took effectively place within a length smaller than $L$. Therefore, while initiation of the buckling at $\tau_b W = (\Delta \theta_b /L) W \approx 0.83$ agreed with what happened for RPBC, the initiation of tube formation at $\tau_t W = (\Delta \theta_t/L)W \approx 0.91$ occured earlier than in RPBC. For increasing length such finite-size effects vanished and the results converged towards RPBC results; for large $L$ buckling and tube-formation points also became sharper. We performed this simulation four times for different twist rates, obtaining invariably the same results.

Such trends in the finite-size effects were confirmed by simulations of shorter ribbons. For instance, dimensional analysis helped to find the scaling $L/W \gtrsim 0.7 \sqrt{W/\text{\AA}}$ as the critical length-to-width ratio above which the picture of the tube-formation process remains valid. Indeed, for $W=25$~\AA, inferring a critical ratio of $\approx 4$, tube was formed in an expected manner for $L/W=7.5$, but not for $L/W=2.6$. More systematic investigations of the finite-size effects are underway.

\section{Concluding discussion}

When the twisted ribbons have atomically smooth edges, which is experimentally feasible and even preferred\cite{cai_nature_10,jia_science_09}, the formed CNTs are expected to become essentially pristine by energy arguments.\cite{malola_PRB_10,koskinen_PRL_08} In practice, however, we cannot exclude the formation of defects either. If tube formation is initiated at different locations with different CNT chiralities\cite{cranford_MSMSE_11}, the zipping-up of the tube may give rise to scattered point defects. Moreover, edge roughness, irregular edge chirality, and edge passivation can lead to CNTs with vacancies, impurity atoms, or dangling bonds, arranged as chiral line defects. Although we cannot entirely exclude the appearance of phenomena related to other finite-size effects\cite{bets_NR_09}, lattice fatigue\cite{nardelli_PRL_98}, or complex defect formation\cite{li_JPD_10}, preliminary results indicate that the central concepts of tube formation prevail. Furthermore, when GNRs are hydrogen-passivated, as they often are, tube formation must be preceded by dehydrogenation and formation of H$_2$ gas. Since this reaction has only a weak thermodynamic driving force\cite{koskinen_PRL_08}, presumable energy barriers for formation of H$_2$ suggest a slow reaction, and catalytic dehydrogenation may be required to aid the tube formation.\cite{shah_EF_04}

To conclude, our study opens up new opportunities in nanomaterial manipulation not limited to carbon-based ribbons alone. Indeed, using a combination of varying geometry that ribbons afford with their separation of scales, one might envisage using inhomogeneous width, chemical modification including passivation, adsorbed  molecules, and clusters, to construct structures with new functionalities. Examples include tubes with bulges or partial tears\cite{tang_PRB_11}, nanoscrolls, multiwalled nanotubes with a spiral cross section\cite{viculis_science_03}, all of which can also be manipulated using external forces so as to enable molecular encapsulation and release in a variety of applications.

\section*{Acknowledgements}
P.K. acknowledges the Academy of Finland and O.K. the National Graduate School of Material Physics (NGSMP) for funding. We acknowledge Karoliina Honkala for discussions, Ville Kotim\"aki for the photo of Fig.~\ref{fig:strap}, and the Finnish IT Center for Science (CSC) for computer resources.

\appendix

\section{About ribbon geometries}
\label{appendix:geometries}
\begin{figure}[h]
\includegraphics[width=6cm]{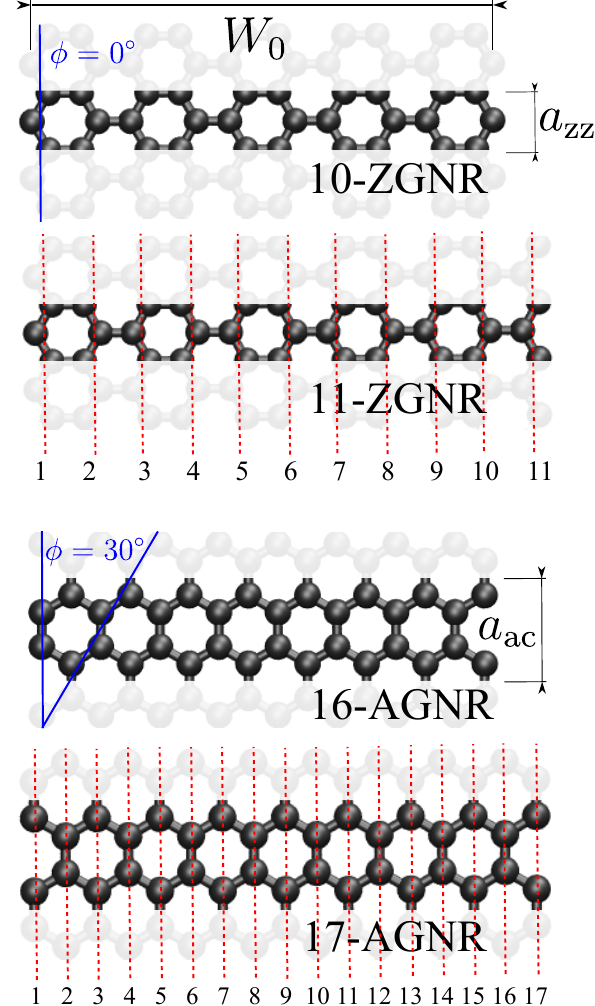}
\caption{(Color online) ZGNRs and AGNRs illustrating the $N$ alternation in the opposite-edge profiles; alternation shows up e.g. in Eq.(\ref{eq:chirality}). The unit-cell length is $a_\text{zz}\approx 2.46$~\AA\ in ZGNRs and $a_\text{ac}\approx 4.26$~\AA\ in AGNRs for zero pre-strain. Ribbon relaxation leads to slightly longer ribbons, more so for narrow ribbons. Width $W_0$ is measured from atomic positions [$W_0\approx(2.13\cdot N - 1.42)$~\AA\ for ZGNRs and $W_0\approx 1.23\cdot(N-1)$~\AA\ for AGNRs]. The angle $\phi$ is the angle between ribbon axis and the adjacent zigzag direction.}
\label{fig:ribbons}
\end{figure}

We simulated tube formation for $N$-ZGNRs ($\phi=0^\circ$, $N=10 - 19$) and $N$-AGNRs ($\phi=30^\circ$, $N=16-34$), their geometries are illustrated in Fig.~\ref{fig:ribbons}. All simulated ZGNRs and AGNRs resulted in pristine CNTs with well-defined chiral indices $(n,m)$ (corresponding to CNTs uniquely defined by the vector ${\bf C}=n {\bf a} + m{\bf b}$, the circumferential vector expressed in terms of the honeycomb unit-cell vectors ${\bf a}$ and ${\bf b}$). For computational feasibility ribbons were unpassivated; hydrogen passivation would have required catalyst particles or a prohibitively long simulation time.


The tubularity parameter was defined as
\begin{equation}
\Omega = \frac{d_\text{flat}-d}{d_\text{flat}-d_\text{bond}},
\label{eq:omega}
\end{equation}
where $d_\text{bond}=1.42$~\AA\ is the carbon-carbon bond length and $d$ is the distance between any two opposite-edge atoms that form a bond in the final tube. The distance is at maximum ($d=d_\text{flat}$) for a flat ribbon and at minimum ($d=d_\text{bond}$) for a tube. Hence $\Omega=0$ for a flat ribbon and $\Omega=1$ for a tube. The threshold for buckling was defined as $\Omega_\text{buckled} > 0.1$. In the continuum limit, because there are no bonds, Eq.(\ref{eq:omega}) was used with $d_\text{bond}=0$.

\begin{figure}[t]
\includegraphics[width=8cm]{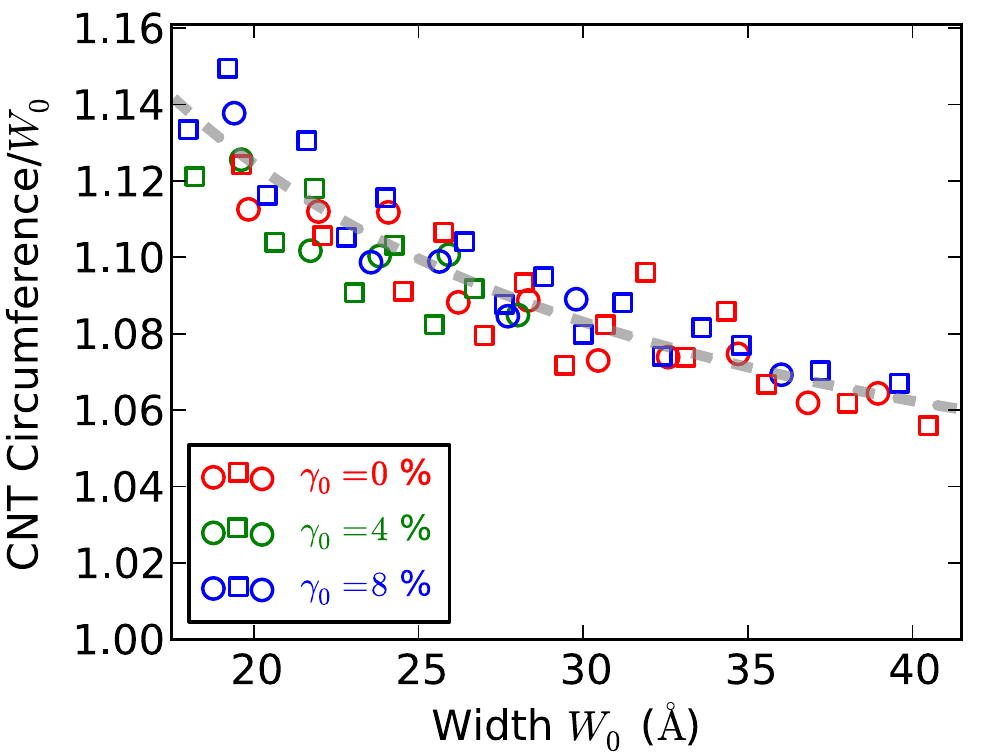}
\caption{(Color online) The ratio of CNT circumference to $W_0$. Fitting gave for the tube circumference $\pi D \equiv W\approx W_0 + 2.5$~\AA\ (Poisson effect caused by the pre-strain was removed). The CNT circumference was obtained more accurately from the chiral indices, $W=\sqrt{3}d_\text{bond}\sqrt{n^2+m^2+mn}$ with $(n,m)$ predicted by Eq.(\ref{eq:chirality}).}
\label{fig:diameter}
\end{figure}

The width of an atomistic ribbon is a question of definition, and although the width $W_0$ (Fig.~\ref{fig:ribbons}) would be an easy concept, the direct comparison of atomic and continuum widths is inherently ambiguous. We chose to define the ribbon width $W$ of the atomistic ribbon as the circumference of the resulting CNT (Fig.~\ref{fig:diameter}). The difference between $W$ and $W_0$, which mainly originates from curvature and bond formation (merging of opposite edges creates `new surface area'), plays a bigger role in narrow ribbons (when $d_\text{bond}$ is a notable fraction of $W_0$).


\section{Molecular-dynamics simulations}
\label{appendix:MD}
We used spin-unpolarized density-functional tight-binding\cite{porezag_PRB_95,koskinen_CMS_09} and revised periodic boundary conditions adapted to chiral symmetry.\cite{dumitrica_JMPS_07,koskinen_PRL_10,koskinen_APL_11,kit_PRB_11} In the twisting simulations, with minimal cell in the axial direction and zero strain corresponding to a relaxed flat ribbon, we used the Langevin thermostat at $300$~K with $1$~fs time step, and a stepped twist rate of $\Delta (\tau W)/\Delta t=0.2-0.3$~ns$^{-1}$. (Buckling and tube formation was possible due to the absence of symmetry constraints with respect to axial symmetry, unlike in Refs \onlinecite{gunlycke_NL_10} and \onlinecite{koskinen_APL_11}.) The rate has only a minor effect on the results because of the abruptness of the buckling and tube-formation events. Molecular-dynamics simulations were performed for ZGNRs using $20$ $\kappa$-points and for AGNRs using $10$ $\kappa$-points (while calculating energy gaps using $100$ $\kappa$-points) with respect to the chiral symmetry operation.

\section{Analysis of twisting of a ribbon based on elasticity theory}
\label{appendix:elasticity}

We consider twisting of a thin ribbon with a fixed length and translational symmetry
such that each cross section of the ribbon has the same shape. A cross section is free to warp in the direction
of the twist axis.
We denote by $\mathbf{x} = (x_1, x_2)$ the material coordinates of the ribbon, where $x_1$ and $x_2$ are the coordinates in the transverse and longitudinal directions, respectively.
Position in space of a material point $\mathbf{x}$ is given by $\mathbf{r}(\mathbf{x}) = (x, y, z)$, and the twist axis was chosen to coincide with the $z$-axis.
Symmetry of the problem implies that
\begin{equation}
\frac{\partial \mathbf{r}}{ \partial x_2} = (-\tau y, \tau x, 1 + \gamma_0)
\end{equation}
 and
\begin{equation}
\frac{\partial^2 \mathbf{r}}{ \partial x_2^2} = (-\tau^2 x, -\tau^2 y, 0),
\end{equation}
where $\tau$ is the twist per unit length and $\gamma_0$ is a longitudinal external strain.
In-plane deformation of the sheet is described by the strain tensor
\begin{equation}
\gamma_{ij} = \frac{1}{2} \left( \frac{ \partial \mathbf{r} }{ \partial x_i} \cdot \frac{ \partial \mathbf{r} }{ \partial x_j} - \delta_{ij} \right)
\end{equation}
and out-of-plane deformation by the curvature tensor
\begin{equation}
C_{ij} = \mathbf{n} \cdot \frac{ \partial^2 \mathbf{r}}{ \partial x_i \partial x_j },
\end{equation}
where $\mathbf{n}$
is the surface normal and $i, j = 1,2$.
Deformation energy per unit length is given by \cite{landau_lifshitz}
\begin{align}
U &= \frac{k}{2(1+\nu)} \int_0^W \left\{ \operatorname{Tr}\boldsymbol\gamma^2 + \frac{\nu}{1-2\nu}(\operatorname{Tr} \boldsymbol\gamma)^2 \right\} dx_1 \\
 &+ \frac{K}{2} \int_0^W \left\{ (\operatorname{Tr} \mathbf{C})^2 + (1-\nu) \left[ \operatorname{Tr} \mathbf{C}^2 - (\operatorname{Tr} \mathbf{C})^2 \right] \right\} dx_1.
\nonumber
\end{align}
Here $k$ is the in-plane modulus, $K$ the bending modulus, $\nu$ the Poisson ratio and $W$ the ribbon width.
The shape of the ribbon was found by numerically minimizing $U$.
To this end we discretized the cross section $x_2 = const$ into $N = 100$ points and replaced the derivatives by finite differences.
The discretized energy with $\nu = 0.3$ was minimized by a damped molecular-dynamics method. Both $\tau_b$ and $\tau_t$ could be determined by increasing $\tau$ in small steps.

\begin{figure*}[t]
\includegraphics[width=12cm]{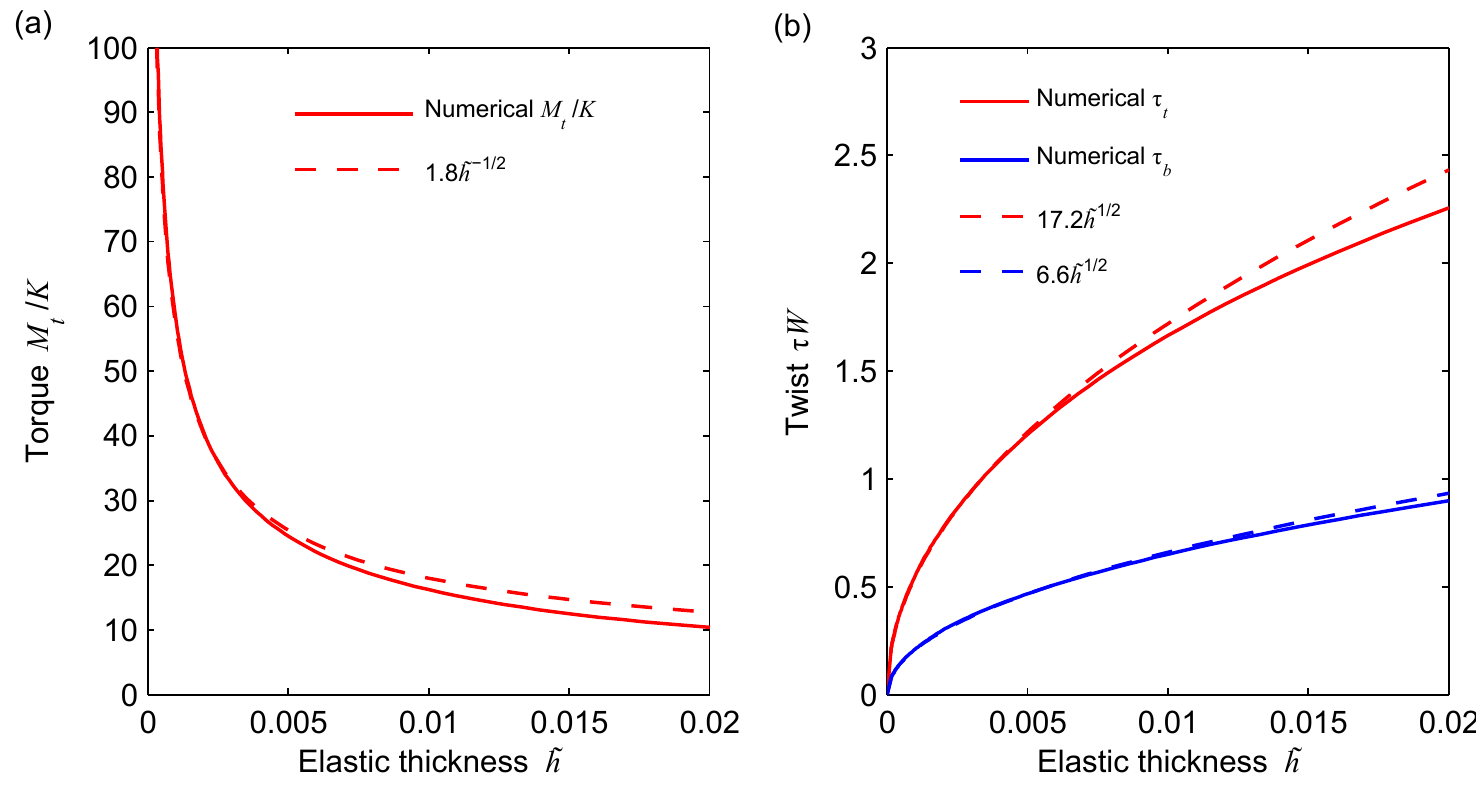}
\caption{(Color online) (a) Torque at the tube-formation point as determined by numerical energy minimization (full line) is compared with the scaling estimate $M_t/K \approx 1.8/\sqrt{\tilde{h}}$ (dashed line). (b) Tube-formation point, $\tau_t$, and buckling point, $\tau_b$, are compared with the scaling estimates $\tau_t \approx 17.2 \sqrt{\tilde{h}}$ and $\tau_b \approx 6.6 \sqrt{\tilde{h}}$, respectively.}
\label{fig:elasticity2}
\end{figure*}

Useful insight into the buckling and tube formation can be obtained by a simple scaling analysis.
During twist the initially straight longitudinal 'fibers' (narrow strips across the ribbon) are deformed into helices and strained by $\frac{1}{2}\tau^2 R^2$, where $R$ is the distance from the twist
axis. This generates a compressive stress $T \sim k \tau^4 W^4$ in the transverse direction of a flat ribbon as the fibers tend towards the axis to minimize the longitudinal stretching energy. The buckling threshold for a compressed plate is given by\cite{landau_lifshitz} $T \sim K/W^2$, from which
a critical twist, $\tau_b W \approx c_b \sqrt{\tilde{h}}$, is obtained. Here $\tilde{h} = \sqrt{K/(kW^2)}$ is the elastic thickness.
For $\tau > \tau_b$ the buckled shape of the cross section is determined by competition between stretching and bending.
The stretching and bending energies per unit length can be estimated such that $S \approx \frac{k}{2}\left( \frac{1}{2}\tau^2 R^2 \right)^2W$ and $B \approx \frac{KW}{2R^2}$, respectively, for a cross section curved with a radius $R$.
By minimizing $S+B$ for $R = \frac{W}{2\pi}$, we find that the twist $\tau_t W \approx c_t \sqrt{\tilde{h}}$ is required to bring the two
edges of the ribbon together so as to form a tube. For
the torque we find that $M_t = \left. \frac{\partial (S+B)}{\partial \tau} \right|_{\tau=\tau_t} \approx c_M K/\sqrt{\tilde{h}}$. The constants $c_b = 6.6$, $c_t = 17.2$, and $c_M = 1.8$ were found by fitting results of numerical energy minimization, see Fig.~\ref{fig:elasticity2}. The simple scaling expressions work well, although deviations appear at high $\tilde{h}$ when the cross-section warps are large.

During twisting nearly all of the shear strain vanishes, which leads to warping of the cross section, i.e., to relative displacement of the two edges along the twist axis. Integrated warp, or the shift $\Delta z$, can be approximated by
\begin{equation} \label{warp}
\Delta z \approx \int_{-W/2}^{W/2} \tau r \left( \hat{\boldsymbol{\Theta}} \cdot \frac{\partial \mathbf{r}}{ \partial x_1 } \right) dx_1,
\end{equation}
where $r$ is the distance from the twist axis and $\hat{\boldsymbol{\Theta}}$ is a unit vector perpendicular to the radial direction.
By assuming a circular cross section with radius $R = \frac{W}{2\pi}$, we find
\begin{equation}
\Delta z \approx \frac{\tau_t W^2}{2\pi}
\end{equation}
for the shift at the tube-formation point.



\begin{thebibliography}{39}
\expandafter\ifx\csname natexlab\endcsname\relax\def\natexlab#1{#1}\fi
\expandafter\ifx\csname bibnamefont\endcsname\relax
  \def\bibnamefont#1{#1}\fi
\expandafter\ifx\csname bibfnamefont\endcsname\relax
  \def\bibfnamefont#1{#1}\fi
\expandafter\ifx\csname citenamefont\endcsname\relax
  \def\citenamefont#1{#1}\fi
\expandafter\ifx\csname url\endcsname\relax
  \def\url#1{\texttt{#1}}\fi
\expandafter\ifx\csname urlprefix\endcsname\relax\def\urlprefix{URL }\fi
\providecommand{\bibinfo}[2]{#2}
\providecommand{\eprint}[2][]{\url{#2}}

\bibitem[{\citenamefont{Iijima}(1991)}]{iijima_nature_91}
\bibinfo{author}{\bibfnamefont{S.}~\bibnamefont{Iijima}},
  \bibinfo{journal}{Nature} \textbf{\bibinfo{volume}{354}}, \bibinfo{pages}{56}
  (\bibinfo{year}{1991}).

\bibitem[{\citenamefont{Saito et~al.}(1998)\citenamefont{Saito, Dresselhaus,
  and Dresselhaus}}]{saito_book_98}
\bibinfo{author}{\bibfnamefont{R.}~\bibnamefont{Saito}},
  \bibinfo{author}{\bibfnamefont{G.}~\bibnamefont{Dresselhaus}},
  \bibnamefont{and} \bibinfo{author}{\bibfnamefont{M.~S.}
  \bibnamefont{Dresselhaus}}, \emph{\bibinfo{title}{Physical Properties of
  Carbon Nanotubes}} (\bibinfo{publisher}{Imperial College Press, London},
  \bibinfo{year}{1998}), \bibinfo{edition}{1st} ed.

\bibitem[{\citenamefont{Kosynkin et~al.}(2009)\citenamefont{Kosynkin,
  Higginbotham, Sinitskii, Lomeda, Dimiev, Prince, and
  Tour}}]{kosynkin_nature_09}
\bibinfo{author}{\bibfnamefont{D.~V.} \bibnamefont{Kosynkin}},
  \bibinfo{author}{\bibfnamefont{A.~L.} \bibnamefont{Higginbotham}},
  \bibinfo{author}{\bibfnamefont{A.}~\bibnamefont{Sinitskii}},
  \bibinfo{author}{\bibfnamefont{J.~R.} \bibnamefont{Lomeda}},
  \bibinfo{author}{\bibfnamefont{A.}~\bibnamefont{Dimiev}},
  \bibinfo{author}{\bibfnamefont{B.~K.} \bibnamefont{Prince}},
  \bibnamefont{and} \bibinfo{author}{\bibfnamefont{J.~M.} \bibnamefont{Tour}},
  \bibinfo{journal}{Nature} \textbf{\bibinfo{volume}{458}},
  \bibinfo{pages}{872} (\bibinfo{year}{2009}).

\bibitem[{\citenamefont{Jiao et~al.}(2009)\citenamefont{Jiao, Zhang, Wang,
  Diankov, and Dai}}]{jiao_nature_09}
\bibinfo{author}{\bibfnamefont{L.}~\bibnamefont{Jiao}},
  \bibinfo{author}{\bibfnamefont{L.}~\bibnamefont{Zhang}},
  \bibinfo{author}{\bibfnamefont{X.}~\bibnamefont{Wang}},
  \bibinfo{author}{\bibfnamefont{G.}~\bibnamefont{Diankov}}, \bibnamefont{and}
  \bibinfo{author}{\bibfnamefont{H.}~\bibnamefont{Dai}},
  \bibinfo{journal}{Nature} \textbf{\bibinfo{volume}{458}},
  \bibinfo{pages}{877} (\bibinfo{year}{2009}).

\bibitem[{\citenamefont{Jia et~al.}(2009)\citenamefont{Jia, Hofmann, Meunier,
  Sumpter, Campos-Delgado, Romo-Herrera, Son, Hsieh, Reina, Kong
  et~al.}}]{jia_science_09}
\bibinfo{author}{\bibfnamefont{X.}~\bibnamefont{Jia}},
  \bibinfo{author}{\bibfnamefont{M.}~\bibnamefont{Hofmann}},
  \bibinfo{author}{\bibfnamefont{V.}~\bibnamefont{Meunier}},
  \bibinfo{author}{\bibfnamefont{B.~G.} \bibnamefont{Sumpter}},
  \bibinfo{author}{\bibfnamefont{J.}~\bibnamefont{Campos-Delgado}},
  \bibinfo{author}{\bibfnamefont{J.~M.} \bibnamefont{Romo-Herrera}},
  \bibinfo{author}{\bibfnamefont{H.}~\bibnamefont{Son}},
  \bibinfo{author}{\bibfnamefont{Y.-P.} \bibnamefont{Hsieh}},
  \bibinfo{author}{\bibfnamefont{A.}~\bibnamefont{Reina}},
  \bibinfo{author}{\bibfnamefont{J.}~\bibnamefont{Kong}}, \bibnamefont{et~al.},
  \bibinfo{journal}{Science} \textbf{\bibinfo{volume}{323}},
  \bibinfo{pages}{1701} (\bibinfo{year}{2009}).

\bibitem[{\citenamefont{Cai et~al.}(2010)\citenamefont{Cai, Ruffieux, Jafaar,
  Bieri, Braun, Blankenburg, Muoth, Seitsonen, Saleh, Feng
  et~al.}}]{cai_nature_10}
\bibinfo{author}{\bibfnamefont{J.}~\bibnamefont{Cai}},
  \bibinfo{author}{\bibfnamefont{P.}~\bibnamefont{Ruffieux}},
  \bibinfo{author}{\bibfnamefont{R.}~\bibnamefont{Jafaar}},
  \bibinfo{author}{\bibfnamefont{M.}~\bibnamefont{Bieri}},
  \bibinfo{author}{\bibfnamefont{T.}~\bibnamefont{Braun}},
  \bibinfo{author}{\bibfnamefont{S.}~\bibnamefont{Blankenburg}},
  \bibinfo{author}{\bibfnamefont{M.}~\bibnamefont{Muoth}},
  \bibinfo{author}{\bibfnamefont{A.~P.} \bibnamefont{Seitsonen}},
  \bibinfo{author}{\bibfnamefont{M.}~\bibnamefont{Saleh}},
  \bibinfo{author}{\bibfnamefont{X.}~\bibnamefont{Feng}}, \bibnamefont{et~al.},
  \bibinfo{journal}{Nature} \textbf{\bibinfo{volume}{466}},
  \bibinfo{pages}{470} (\bibinfo{year}{2010}).

\bibitem[{\citenamefont{Hamada et~al.}(1992)\citenamefont{Hamada, Sawada, and
  Oshiyama}}]{hamada_PRL_92}
\bibinfo{author}{\bibfnamefont{N.}~\bibnamefont{Hamada}},
  \bibinfo{author}{\bibfnamefont{S.}~\bibnamefont{Sawada}}, \bibnamefont{and}
  \bibinfo{author}{\bibfnamefont{A.}~\bibnamefont{Oshiyama}},
  \bibinfo{journal}{Phys. Rev. Lett.} \textbf{\bibinfo{volume}{68}},
  \bibinfo{pages}{1579} (\bibinfo{year}{1992}).

\bibitem[{\citenamefont{Fennimore et~al.}(2003)\citenamefont{Fennimore,
  Yuzvinsky, Han, Fuhrer, Cumings, and Zettl}}]{fennimore_nature_03}
\bibinfo{author}{\bibfnamefont{A.~M.} \bibnamefont{Fennimore}},
  \bibinfo{author}{\bibfnamefont{T.~D.} \bibnamefont{Yuzvinsky}},
  \bibinfo{author}{\bibfnamefont{W.-Q.} \bibnamefont{Han}},
  \bibinfo{author}{\bibfnamefont{M.~S.} \bibnamefont{Fuhrer}},
  \bibinfo{author}{\bibfnamefont{J.}~\bibnamefont{Cumings}}, \bibnamefont{and}
  \bibinfo{author}{\bibfnamefont{A.}~\bibnamefont{Zettl}},
  \bibinfo{journal}{Nature} \textbf{\bibinfo{volume}{424}},
  \bibinfo{pages}{408} (\bibinfo{year}{2003}).

\bibitem[{\citenamefont{Meyer et~al.}(2005)\citenamefont{Meyer, Paillet, and
  Roth}}]{meyer_science_05}
\bibinfo{author}{\bibfnamefont{J.~C.} \bibnamefont{Meyer}},
  \bibinfo{author}{\bibfnamefont{M.}~\bibnamefont{Paillet}}, \bibnamefont{and}
  \bibinfo{author}{\bibfnamefont{S.}~\bibnamefont{Roth}},
  \bibinfo{journal}{Science} \textbf{\bibinfo{volume}{309}},
  \bibinfo{pages}{1539} (\bibinfo{year}{2005}).

\bibitem[{\citenamefont{Green}(1936)}]{green_PRSL_36}
\bibinfo{author}{\bibfnamefont{A.~E.} \bibnamefont{Green}},
  \bibinfo{journal}{Proc. R. Soc. London} \textbf{\bibinfo{volume}{154}},
  \bibinfo{pages}{430} (\bibinfo{year}{1936}).

\bibitem[{\citenamefont{Green}(1937)}]{green_PRSL_37}
\bibinfo{author}{\bibfnamefont{A.~E.} \bibnamefont{Green}},
  \bibinfo{journal}{Proc. R. Soc. London} \textbf{\bibinfo{volume}{161}},
  \bibinfo{pages}{197} (\bibinfo{year}{1937}).

\bibitem[{\citenamefont{Mansfield}(1989)}]{mansfield_book_89}
\bibinfo{author}{\bibfnamefont{E.~H.} \bibnamefont{Mansfield}},
  \emph{\bibinfo{title}{The Bending and Stretching of Plates}}
  (\bibinfo{publisher}{Cambridge University Press, Cambridge},
  \bibinfo{year}{1989}), \bibinfo{edition}{2nd} ed.

\bibitem[{\citenamefont{Korte et~al.}(2010)\citenamefont{Korte, Starostin, and
  {van der Heijden}}}]{korte_PRSL_10}
\bibinfo{author}{\bibfnamefont{A.~P.} \bibnamefont{Korte}},
  \bibinfo{author}{\bibfnamefont{E.~L.} \bibnamefont{Starostin}},
  \bibnamefont{and} \bibinfo{author}{\bibfnamefont{G.~H.~M.} \bibnamefont{{van
  der Heijden}}}, \bibinfo{journal}{Proceedings of the Royal Society of Longon.
  Series A} \textbf{\bibinfo{volume}{47}}, \bibinfo{pages}{285}
  (\bibinfo{year}{2010}).

\bibitem[{\citenamefont{Koskinen and
  Kit}(2010{\natexlab{a}})}]{koskinen_PRB_10b}
\bibinfo{author}{\bibfnamefont{P.}~\bibnamefont{Koskinen}} \bibnamefont{and}
  \bibinfo{author}{\bibfnamefont{O.~O.} \bibnamefont{Kit}},
  \bibinfo{journal}{Phys. Rev. B} \textbf{\bibinfo{volume}{81}},
  \bibinfo{pages}{235420} (\bibinfo{year}{2010}{\natexlab{a}}).

\bibitem[{\citenamefont{Liang and Mahadevan}(2009)}]{liang_PNAS_09}
\bibinfo{author}{\bibfnamefont{H.}~\bibnamefont{Liang}} \bibnamefont{and}
  \bibinfo{author}{\bibfnamefont{L.}~\bibnamefont{Mahadevan}},
  \bibinfo{journal}{Proc. Nat. Ac. Sci.} \textbf{\bibinfo{volume}{106}},
  \bibinfo{pages}{22049} (\bibinfo{year}{2009}).

\bibitem[{\citenamefont{Huang et~al.}(2009)\citenamefont{Huang, Liu, Su, Wu,
  Duan, Gu, and Liu}}]{bhuang_PRL_09}
\bibinfo{author}{\bibfnamefont{B.}~\bibnamefont{Huang}},
  \bibinfo{author}{\bibfnamefont{M.}~\bibnamefont{Liu}},
  \bibinfo{author}{\bibfnamefont{N.}~\bibnamefont{Su}},
  \bibinfo{author}{\bibfnamefont{J.}~\bibnamefont{Wu}},
  \bibinfo{author}{\bibfnamefont{W.}~\bibnamefont{Duan}},
  \bibinfo{author}{\bibfnamefont{B.}~\bibnamefont{Gu}}, \bibnamefont{and}
  \bibinfo{author}{\bibfnamefont{F.}~\bibnamefont{Liu}},
  \bibinfo{journal}{Phys. Rev. Lett.} \textbf{\bibinfo{volume}{102}},
  \bibinfo{pages}{166404} (\bibinfo{year}{2009}).

\bibitem[{\citenamefont{Barone et~al.}(2006)\citenamefont{Barone, Hod, and
  Scuseria}}]{barone_NL_06}
\bibinfo{author}{\bibfnamefont{V.}~\bibnamefont{Barone}},
  \bibinfo{author}{\bibfnamefont{O.}~\bibnamefont{Hod}}, \bibnamefont{and}
  \bibinfo{author}{\bibfnamefont{G.~E.} \bibnamefont{Scuseria}},
  \bibinfo{journal}{Nano Lett.} \textbf{\bibinfo{volume}{6}},
  \bibinfo{pages}{2748} (\bibinfo{year}{2006}).

\bibitem[{\citenamefont{Koskinen and
  Kit}(2010{\natexlab{b}})}]{koskinen_PRL_10}
\bibinfo{author}{\bibfnamefont{P.}~\bibnamefont{Koskinen}} \bibnamefont{and}
  \bibinfo{author}{\bibfnamefont{O.~O.} \bibnamefont{Kit}},
  \bibinfo{journal}{Phys. Rev. Lett.} \textbf{\bibinfo{volume}{105}},
  \bibinfo{pages}{106401} (\bibinfo{year}{2010}{\natexlab{b}}).

\bibitem[{\citenamefont{Gunlycke et~al.}(2010)\citenamefont{Gunlycke, Li,
  Mintmire, and White}}]{gunlycke_NL_10}
\bibinfo{author}{\bibfnamefont{D.}~\bibnamefont{Gunlycke}},
  \bibinfo{author}{\bibfnamefont{J.}~\bibnamefont{Li}},
  \bibinfo{author}{\bibfnamefont{J.~W.} \bibnamefont{Mintmire}},
  \bibnamefont{and} \bibinfo{author}{\bibfnamefont{C.~T.} \bibnamefont{White}},
  \bibinfo{journal}{Nano Lett.} \textbf{\bibinfo{volume}{10}},
  \bibinfo{pages}{3638} (\bibinfo{year}{2010}).

\bibitem[{\citenamefont{Zhang and Dumitric{\u a}}(2011)}]{zhang_small_11}
\bibinfo{author}{\bibfnamefont{D.-B.} \bibnamefont{Zhang}} \bibnamefont{and}
  \bibinfo{author}{\bibfnamefont{T.}~\bibnamefont{Dumitric{\u a}}},
  \bibinfo{journal}{Small} \textbf{\bibinfo{volume}{7}}, \bibinfo{pages}{1023}
  (\bibinfo{year}{2011}).

\bibitem[{\citenamefont{Koskinen}(2011)}]{koskinen_APL_11}
\bibinfo{author}{\bibfnamefont{P.}~\bibnamefont{Koskinen}},
  \bibinfo{journal}{Appl. Phys. Lett.} \textbf{\bibinfo{volume}{99}},
  \bibinfo{pages}{013105} (\bibinfo{year}{2011}).

\bibitem[{\citenamefont{Popov}(2004)}]{popov_NJP_04}
\bibinfo{author}{\bibfnamefont{V.~N.} \bibnamefont{Popov}},
  \bibinfo{journal}{New J. Phys.} \textbf{\bibinfo{volume}{6}},
  \bibinfo{pages}{17} (\bibinfo{year}{2004}).

\bibitem[{\citenamefont{Plimpton}(1995)}]{plimpton_JCP_95}
\bibinfo{author}{\bibfnamefont{S.}~\bibnamefont{Plimpton}},
  \bibinfo{journal}{J. Comp. Phys.} \textbf{\bibinfo{volume}{117}},
  \bibinfo{pages}{1} (\bibinfo{year}{1995}).

\bibitem[{lam()}]{lammps}
\bibinfo{note}{\texttt{http://lammps.sandia.gov}}.

\bibitem[{\citenamefont{Brenner et~al.}(2002)\citenamefont{Brenner, Shenderova,
  Harrison, Stuart, Ni, and Sinnott}}]{brenner_JPCM_02}
\bibinfo{author}{\bibfnamefont{D.~W.} \bibnamefont{Brenner}},
  \bibinfo{author}{\bibfnamefont{O.~A.} \bibnamefont{Shenderova}},
  \bibinfo{author}{\bibfnamefont{J.~A.} \bibnamefont{Harrison}},
  \bibinfo{author}{\bibfnamefont{S.~J.} \bibnamefont{Stuart}},
  \bibinfo{author}{\bibfnamefont{B.}~\bibnamefont{Ni}}, \bibnamefont{and}
  \bibinfo{author}{\bibfnamefont{S.~B.} \bibnamefont{Sinnott}},
  \bibinfo{journal}{J. Phys.: Condens. Matter} \textbf{\bibinfo{volume}{14}},
  \bibinfo{pages}{783} (\bibinfo{year}{2002}).

\bibitem[{\citenamefont{Malola et~al.}(2010)\citenamefont{Malola, H\"akkinen,
  and Koskinen}}]{malola_PRB_10}
\bibinfo{author}{\bibfnamefont{S.}~\bibnamefont{Malola}},
  \bibinfo{author}{\bibfnamefont{H.}~\bibnamefont{H\"akkinen}},
  \bibnamefont{and} \bibinfo{author}{\bibfnamefont{P.}~\bibnamefont{Koskinen}},
  \bibinfo{journal}{Phys. Rev. B} \textbf{\bibinfo{volume}{81}},
  \bibinfo{pages}{165447} (\bibinfo{year}{2010}).

\bibitem[{\citenamefont{Koskinen et~al.}(2008)\citenamefont{Koskinen, Malola,
  and H\"akkinen}}]{koskinen_PRL_08}
\bibinfo{author}{\bibfnamefont{P.}~\bibnamefont{Koskinen}},
  \bibinfo{author}{\bibfnamefont{S.}~\bibnamefont{Malola}}, \bibnamefont{and}
  \bibinfo{author}{\bibfnamefont{H.}~\bibnamefont{H\"akkinen}},
  \bibinfo{journal}{Phys. Rev. Lett.} \textbf{\bibinfo{volume}{101}},
  \bibinfo{pages}{115502} (\bibinfo{year}{2008}).

\bibitem[{\citenamefont{Cranford and Buehler}(2011)}]{cranford_MSMSE_11}
\bibinfo{author}{\bibfnamefont{S.}~\bibnamefont{Cranford}} \bibnamefont{and}
  \bibinfo{author}{\bibfnamefont{M.~J.} \bibnamefont{Buehler}},
  \bibinfo{journal}{Modelling Simul. Mater. Sci. Eng.}
  \textbf{\bibinfo{volume}{19}}, \bibinfo{pages}{054003}
  (\bibinfo{year}{2011}).

\bibitem[{\citenamefont{Bets and Yakobson}(2009)}]{bets_NR_09}
\bibinfo{author}{\bibfnamefont{K.~V.} \bibnamefont{Bets}} \bibnamefont{and}
  \bibinfo{author}{\bibfnamefont{B.~I.} \bibnamefont{Yakobson}},
  \bibinfo{journal}{Nano Res} \textbf{\bibinfo{volume}{2}},
  \bibinfo{pages}{161} (\bibinfo{year}{2009}).

\bibitem[{\citenamefont{Nardelli et~al.}(1998)\citenamefont{Nardelli, Yakobson,
  and Bernholc}}]{nardelli_PRL_98}
\bibinfo{author}{\bibfnamefont{M.~B.} \bibnamefont{Nardelli}},
  \bibinfo{author}{\bibfnamefont{B.~I.} \bibnamefont{Yakobson}},
  \bibnamefont{and} \bibinfo{author}{\bibfnamefont{J.}~\bibnamefont{Bernholc}},
  \bibinfo{journal}{Phys. Rev. Lett.} \textbf{\bibinfo{volume}{81}},
  \bibinfo{pages}{4656} (\bibinfo{year}{1998}).

\bibitem[{\citenamefont{Li}(2010)}]{li_JPD_10}
\bibinfo{author}{\bibfnamefont{Y.}~\bibnamefont{Li}}, \bibinfo{journal}{J.
  Phys. D: Appl. Phys.} \textbf{\bibinfo{volume}{43}}, \bibinfo{pages}{495405}
  (\bibinfo{year}{2010}).

\bibitem[{\citenamefont{Shah et~al.}(2004)\citenamefont{Shah, Wang, Panjala,
  and Huffman}}]{shah_EF_04}
\bibinfo{author}{\bibfnamefont{N.}~\bibnamefont{Shah}},
  \bibinfo{author}{\bibfnamefont{Y.}~\bibnamefont{Wang}},
  \bibinfo{author}{\bibfnamefont{D.}~\bibnamefont{Panjala}}, \bibnamefont{and}
  \bibinfo{author}{\bibfnamefont{G.~P.} \bibnamefont{Huffman}},
  \bibinfo{journal}{Energy \& Fuels} \textbf{\bibinfo{volume}{18}},
  \bibinfo{pages}{727} (\bibinfo{year}{2004}).

\bibitem[{\citenamefont{Tang et~al.}(2011)\citenamefont{Tang, Guo, and
  Chen}}]{tang_PRB_11}
\bibinfo{author}{\bibfnamefont{C.}~\bibnamefont{Tang}},
  \bibinfo{author}{\bibfnamefont{W.}~\bibnamefont{Guo}}, \bibnamefont{and}
  \bibinfo{author}{\bibfnamefont{C.}~\bibnamefont{Chen}},
  \bibinfo{journal}{Phys. Rev. B} \textbf{\bibinfo{volume}{83}},
  \bibinfo{pages}{075410} (\bibinfo{year}{2011}).

\bibitem[{\citenamefont{Viculis et~al.}(2003)\citenamefont{Viculis, Mack, and
  Kaner}}]{viculis_science_03}
\bibinfo{author}{\bibfnamefont{L.~M.} \bibnamefont{Viculis}},
  \bibinfo{author}{\bibfnamefont{J.~J.} \bibnamefont{Mack}}, \bibnamefont{and}
  \bibinfo{author}{\bibfnamefont{R.~B.} \bibnamefont{Kaner}},
  \bibinfo{journal}{Science} \textbf{\bibinfo{volume}{299}},
  \bibinfo{pages}{1361} (\bibinfo{year}{2003}).

\bibitem[{\citenamefont{Porezag et~al.}(1995)\citenamefont{Porezag, Frauenheim,
  K\"ohler, Seifert, and Kaschner}}]{porezag_PRB_95}
\bibinfo{author}{\bibfnamefont{D.}~\bibnamefont{Porezag}},
  \bibinfo{author}{\bibfnamefont{T.}~\bibnamefont{Frauenheim}},
  \bibinfo{author}{\bibfnamefont{T.}~\bibnamefont{K\"ohler}},
  \bibinfo{author}{\bibfnamefont{G.}~\bibnamefont{Seifert}}, \bibnamefont{and}
  \bibinfo{author}{\bibfnamefont{R.}~\bibnamefont{Kaschner}},
  \bibinfo{journal}{Phys. Rev. B} \textbf{\bibinfo{volume}{51}},
  \bibinfo{pages}{12947} (\bibinfo{year}{1995}).

\bibitem[{\citenamefont{Koskinen and M\"akinen}(2009)}]{koskinen_CMS_09}
\bibinfo{author}{\bibfnamefont{P.}~\bibnamefont{Koskinen}} \bibnamefont{and}
  \bibinfo{author}{\bibfnamefont{V.}~\bibnamefont{M\"akinen}},
  \bibinfo{journal}{Comput. Mater. Sci.} \textbf{\bibinfo{volume}{47}},
  \bibinfo{pages}{237} (\bibinfo{year}{2009}).

\bibitem[{\citenamefont{Dumitric{\u a} and James}(2007)}]{dumitrica_JMPS_07}
\bibinfo{author}{\bibfnamefont{T.}~\bibnamefont{Dumitric{\u a}}}
  \bibnamefont{and} \bibinfo{author}{\bibfnamefont{R.~D.} \bibnamefont{James}},
  \bibinfo{journal}{J. Mech. Phys. Solids} \textbf{\bibinfo{volume}{55}},
  \bibinfo{pages}{2206} (\bibinfo{year}{2007}).

\bibitem[{\citenamefont{Kit et~al.}(2011)\citenamefont{Kit, Pastewka, and
  Koskinen}}]{kit_PRB_11}
\bibinfo{author}{\bibfnamefont{O.~O.} \bibnamefont{Kit}},
  \bibinfo{author}{\bibfnamefont{L.}~\bibnamefont{Pastewka}}, \bibnamefont{and}
  \bibinfo{author}{\bibfnamefont{P.}~\bibnamefont{Koskinen}},
  \bibinfo{journal}{Phys. Rev. B} \textbf{\bibinfo{volume}{84}},
  \bibinfo{pages}{155431} (\bibinfo{year}{2011}).

\bibitem[{\citenamefont{Landau and Lifshitz}(1986)}]{landau_lifshitz}
\bibinfo{author}{\bibfnamefont{L.~D.} \bibnamefont{Landau}} \bibnamefont{and}
  \bibinfo{author}{\bibfnamefont{E.~M.} \bibnamefont{Lifshitz}},
  \emph{\bibinfo{title}{Theory of elasticity}} (\bibinfo{publisher}{Pergamon,
  New York}, \bibinfo{year}{1986}), \bibinfo{edition}{3rd} ed.

\end{thebibliography}

\end{document}